\documentclass[aps,pra,twocolumn,amsmath,amssymb,showpacs]{revtex4}
\usepackage{graphics}
\usepackage{epsfig}

\begin{document}
\title{Genuine tripartite entanglement in the non-interacting Fermi gas}
\author{T. V\'ertesi}
\affiliation{Institute of Nuclear Research of the Hungarian
Academy of Sciences,\\ H-4001 Debrecen, P.O. Box 51, Hungary}

\date{\today}

\begin{abstract}
We study genuine tripartite entanglement shared among the spins of
three localized fermions in the non-interacting Fermi gas at zero
temperature. Firstly, we prove analytically with the aid of
entanglement witnesses that in a particular configuration the
three fermions are genuinely tripartite entangled. Then various
three-fermion configurations are investigated in order to quantify
and calculate numerically the amount of genuine tripartite
entanglement present in the system. Further we give a lower and an
upper limit to the maximum diameter of the three-fermion
configuration below which genuine tripartite entanglement exists
and find that this distance is comparable with the maximum
separation between two entangled fermions. The upper and lower
limit turn to be very close to each other indicating that the
applied witness operator is well suited to reveal genuine
tripartite entanglement in the collection of non-interacting
fermions.
\end{abstract}


\maketitle

\section{Introduction}
\label{sec:intro}

Entanglement is in the heart of quantum mechanics and of great
importance in quantum information theory. Two entangled particles
already offer a valuable resource to perform several practical
tasks such as quantum teleportation, quantum cryptography or
quantum computation \cite{NC}. However, the multipartite setting
due to the much richer structure suggests many new possibilities
and phenomena over the bipartite case. Indeed, multipartite states
may contradict local realistic models in a qualitatively different
and stronger way \cite{GHSZ}. Moreover, this feature allows to
implement novel quantum information processing tasks such as
quantum computation based on cluster states \cite{RB},
entanglement enhanced measurements \cite{GLM}, quantum
communication without a common reference frame \cite{BRS} and
open-destination teleportation \cite{Zhao04}.

Though the characterization of multipartite entanglement is
studied in great depth \cite{Acin,PV}, still rarely investigated
in solid state systems (see \cite{GTB} and references therein),
however, it definitely plays an essential role in quantum phase
transitions~\cite{ON02,Oli06} and might well be a key ingredient
to unresolved problems in physics such as high temperature
superconductivity~\cite{Ved04}. In this article we investigate
genuine multipartite entanglement shared among the spins of three
fermions in the Fermi gas of non-interacting particles at zero
temperature (degenerate Fermi gas) following the work of
Refs.~\cite{Ved03,LBV}. We apply an entanglement witness developed
in Ref.~\cite{GTB} (and also appeared in Ref.~\cite{TG}) in order
to reveal genuine tripartite quantum correlations in the
collection of fermions and using the approach in Ref.~\cite{Bra05}
we also characterize quantitatively the amount of it.

The article is organized as follows: In Sec.~\ref{sec:analysis}
according to Refs. \cite{Ved03,LBV} we present the three-spin
reduced density matrix of the degenerate Fermi gas, and show by
analytical means (extending the related results of
Ref.~\cite{LBV}) that both $GHZ$-type and $W$-type witnesses are
unable to detect genuine tripartite entanglement (GTE) among the
spins of three localized fermions. On the other, we demonstrate
that for specific configurations of the three fermions the GTE
witness of Ref.~\cite{GTB} is capable to signal GTE both for the
two- and three-dimensional (2D and 3D) degenerate Fermi gases. In
Sec.~\ref{sec:er} on the basis of this witness a formula is
constructed to the lower bound of the generalized robustness
($E_R$) of genuine tripartite entanglement. With the aid of this
formula in Sec.~\ref{sec:num} we quantify numerically genuine
tripartite quantum correlations for various arrangements of the
three particles. In Sec.~\ref{sec:gte} we determine lower and
upper bounds to the GTE distance (i.e., to the largest diameter of
the three-fermion configuration below which GTE is still present
in the system) both in the 2D and 3D degenerate Fermi gases. The
paper concludes in Sec.~\ref{sec:disc} with a brief summary of the
results obtained and discusses possible schemes to extract GTE
from the system.

\section{Analysis of the density matrix for three fermions}
\label{sec:analysis}
\subsection{Three-spin reduced density matrix}
\label{subsec:3spin}

Consider a system of $N$ non-interacting fermions in a box with
volume $V$. At zero temperature the ground state of the system is
$|\phi_0\rangle = \Pi_k^{k_F}
\hat{c}_{k,\sigma}^\dagger|vac\rangle$, where $k_F = (3\pi^2
N/V)^{1/3}$ is the Fermi momentum and $|vac\rangle$ denotes the
vacuum state. From the ground state of the system $|\phi_0\rangle$
one obtains the three-spin reduced density matrix (up to
normalization) between the three fermions localized at positions
$\mathbf{r}$, $\mathbf{r}'$, and $\mathbf{r}''$,
\begin{align}
&\rho_3(s,s',s'';t,t',t'') = \nonumber \\
&\langle\phi_0|
\hat{\psi}_{t}^\dagger(\mathbf{r})
\hat{\psi}_{t'}^\dagger(\mathbf{r}')
\hat{\psi}_{t''}^\dagger(\mathbf{r}'')
\hat{\psi}_{s''}(\mathbf{r}'') \hat{\psi}_{s'}(\mathbf{r}')
\hat{\psi}_{s}(\mathbf{r}) |\phi_0\rangle
 \;,
 \label{rho3gen}
\end{align}
where $\hat{\psi}_{s}(\mathbf{r})$,
$\hat{\psi}_{s}^\dagger(\mathbf{r})$ are field
annihilation/creation operators for a particle with spin $s$
located at position $\mathbf{r}$ satisfying $\{
\hat{\psi}_{s}(\mathbf{r}),\hat{\psi}_{s}^\dagger(\mathbf{r})\}=\delta_{s,s'}\delta(\mathbf
r-\mathbf r')$. The justification, that the above matrix elements
indeed describe three-qubit quantum states is discussed in the
Appendix of Ref~\cite{Cav05}. Following Refs.~\cite{Ved03,LBV} the
explicit formula for the three-spin reduced density matrix
$\rho_3$ is given by
\begin{align}
\rho_3=(1-p)\frac{\mathbb I_8}{8} &+p_{12}|\Psi_{12}^-\rangle
\langle \Psi_{12}^-|\otimes\frac{\mathbb I_4}{2} +
p_{13}|\Psi_{13}^-\rangle \langle \Psi_{13}^-|\otimes\frac{\mathbb
I_4}{2} \nonumber \\
& +p_{23}|\Psi_{23}^-\rangle \langle
\Psi_{23}^-|\otimes\frac{\mathbb I_4}{2} \;, \label{rho3}
\end{align}
where $p=p_{12}+p_{13}+p_{23}$ and $\mathbb I_n$ denotes the
$n\times n$ identity matrix. Further,
$\Psi_{ij}^-=(|\uparrow\downarrow\rangle-|\downarrow\uparrow\rangle)/\sqrt
2$ is the singlet state of the pair $ij$ in the orthonormal basis
$\{|\uparrow\rangle,|\downarrow\rangle\}$. The value $p_{ij}$
depends only on the relative distance between the three fermions
and can be written explicitly for the fermion pair $ij$ as
\cite{LBV}
\begin{equation}
p_{ij}=\frac{-f_{ij}^2+f_{ij}f_{ik}f_{jk}}{-2+f_{ij}^2+f_{ik}^2+f_{jk}^2-f_{ij}f_{ik}f_{jk}}\;,
\label{pij}
\end{equation}
where the analytic form of $f_{ij}$ depends on the spatial
dimension of the system, that is we may write
\begin{align}
f^{\mathrm{2D}}_{ij} &= 2J_1(k_F r_{ij})/k_F r_{ij} \nonumber \\
f^{\mathrm{3D}}_{ij} &= 3j_1(k_F r_{ij})/k_F r_{ij}
 \label{fij}
\end{align}
in the case of the two- and three-dimensional Fermi gases
\cite{OK}. In the above formulae $j_1$ and $J_1$ denote the
spherical and the first order Bessel function of the first kind,
respectively.

Actually, owing to the collective $SU(2)$ rotational symmetry of
the model Hamiltonian of non-interacting fermions, many matrix
elements of $\rho_3$ in (\ref{rho3}) are forced to be zero.
Explicitly, the states which are invariant under collective
$SU(2)$ rotation of the three qubits are the three-qubit Werner
states \cite{EW}, and they can be given in the form \cite{EW}
\begin{equation}
\rho=\sum_{k=+,0,1,2,3}{\frac{r_k}{4}R_k}\;, \label{rho}
\end{equation}
where $R_k$ are certain linear combinations of permutation
operators and $r_k(\rho)=\mathrm{Tr}(\rho R_k)$. Using the
definitions for $R_k$ from Ref.~\cite{EW} and the explicit form of
the state $\rho_3$ from (\ref{rho3}), we are able to calculate the
parameters $r_k$ for the three-spin reduced density matrix
$\rho_3$, which read as follows
\begin{align}
r_+ & =\frac{1-p}{2} \nonumber \\
r_0 & =\frac{1+p}{2} \nonumber \\
r_1 & =\frac{p_{12}+p_{13}-2p_{23}}{2} \nonumber \\
r_2 & =\frac{3}{2\sqrt 3}(p_{13}-p_{12}) \nonumber \\
r_3 & =0 \;. \label{rk}
\end{align}

\subsection{Possible range of parameters $p_{ij}$}
\label{subsec:param}

According to the Lemma 2 of Ref.~\cite{EW} $\rho$ in (\ref{rho})
is a density matrix only if $r_+,r_0\geq 0$. These inequalities
imply for the state $\rho_3$ by the virtue of (\ref{rk}) that $p$
lies in the interval
\begin{equation}
-1\leq p\leq +1\;. \label{p}
\end{equation}
Let us observe in (\ref{fij}) that $|f_{ij}|\leq 1$ both for the
2D and 3D Fermi gases. This fact together with the bound to $p$ in
(\ref{p}) and also the definition $p=p_{12}+p_{13}+p_{23}$, after
some algebraic manipulations (which are not detailed here), lead
to the bounds
\begin{equation}
-1\leq p_{ij}\leq 1 \label{pijbound}
\end{equation}
for the three different fermion pairs $ij=12,13,23$. Thus, the
parameters $p_{ij}$  appearing in state $\rho_3$ are limited by
the values $\pm1$.

Let us introduce the class of biseparable three-qubit states $B$,
i.e., the states
\begin{equation}
\rho = \sum_i{p_i |\psi_i\rangle\langle \psi_i|}\;, \label{bisep}
\end{equation}
which can be expressed as a convex sum of projectors onto product
and bipartite entangled vectors \cite{Acin}. In
definition~(\ref{bisep}) the pure states $|\psi_i\rangle$ are
separable on the Hilbert-space of three qubits $1,2,3$ with
respect to one of the three bipartitions $1|23$, $12|3$ or $13|2$
and $p_i\geq 0$ adding up to $1$. We say that a general
three-qubit state is genuine tripartite entangled when it is not
in the class of biseparable states $B$, that is they cannot be
constructed by mixing pure states containing bipartite
entanglement at most. Clearly, if $p_{ij}$ in (\ref{rho3}) was
positive for each of the three different pairs (whose sum $p$ is
upper bounded by $+1$ according to~(\ref{p})), then $\rho_3$
should define a biseparable state.

Note, that the bound to $p_{ij}$'s in (\ref{pijbound}) may not be
tight, therefore it is not evident whether $p_{ij}$ can take up
negative values at all. However, by arranging the three fermions
in a particular geometry, we demonstrate that $p_{ij}$ may take
the value $-1/3$ as well: Taking the first and second derivatives
of the functions $f_{ij}(k_F r)$ (defined for both the 2D and 3D
Fermi gases under equations~(\ref{fij})) with respect to $x=k_F
r$, we observe that in the limit $x\rightarrow 0$ they behave as
\begin{equation}
\lim_{x\rightarrow 0}f_{ij}(x)=1,\;\; \lim_{x\rightarrow
0}f'_{ij}(x)=0,\;\; \lim_{x\rightarrow 0}f''_{ij}(x)\neq0 \;.
\label{limits}
\end{equation}
Now let us place the three particles on a line so that particle
$2$ would lie just at the midpoint between particle $1$ and
particle $3$, and let the relative distance between these outer
particles tends to zero. Actually, in the limit $x\rightarrow 0$,
$p_{ij}$ in~(\ref{pij}) can be given explicitly by applying
l'Hospital's rule twice and by taking account the limiting
values~(\ref{limits}). As a result we obtain
\begin{equation}
p_{12}=p_{23}=2/3,\hspace{0.5cm}p_{13}=-1/3 \label{plimit}
\end{equation}
both for the 2D and 3D Fermi gases.

In the next two subsections we propose witness operators in order
to reveal GTE in the degenerate Fermi systems. An observable which
is well suited for signaling genuine tripartite quantum
correlations in Heisenberg spin lattices \cite{GTB} turns out to
detect GTE in the degenerate Fermi gas as well.

\subsection{Generalized $GHZ$ and $W$ witnesess}
\label{subsec:ghzw}

For deciding whether the state $\rho_3$ with the explicit
parameters $p_{12}=p_{23}=2/3$, $p_{13}=-1/3$ in (\ref{plimit}) is
genuine tripartite entangled, we will use entanglement witnesses.
A witness of genuine tripartite entanglement is an observable
$\Pi$ with a positive mean value on all biseparable states so a
negative expectation value $\mathrm{Tr}(\rho \Pi)$ guarantees that
the state $\rho$ carries genuine tripartite entanglement
\cite{Terhal}. Thus a witness operator which separates genuine
tripartite entangled states from the biseparable set $B$ (defined
by equation~(\ref{bisep})) can be given in the form \cite{Bou04}
\begin{equation}
W_\psi = \Lambda \mathbb I_8 - |\psi\rangle\langle\psi|\;,
\label{wit}
\end{equation}
where
\begin{equation}
\Lambda = \max_{|\phi\rangle \in B} |\langle
\phi|\psi\rangle|^2\;.
\end{equation}

A simple method has been found in Ref.~\cite{Bou04} to determine
$\Lambda$ for any pure genuine tripartite entangled state
$|\psi\rangle$. In particular, let $|\psi\rangle$ be the
$GHZ$-like and the $W$-like state \cite{CC05}, which are
respectively
\begin{align}
|GHZ(\alpha)\rangle &= \frac{|\mathbf n_1,\mathbf n_2, \mathbf
n_3\rangle+
e^{i\alpha}|-\mathbf n_1,-\mathbf n_2,-\mathbf n_3\rangle}{\sqrt 2} \nonumber \\
|W(\beta,\gamma)\rangle &= \frac{|\mathbf n_1,\mathbf n_2,-\mathbf
n_3\rangle +e^{i\beta}|\mathbf n_1,-\mathbf n_2,\mathbf
n_3\rangle+e^{i\gamma} |-\mathbf n_1,\mathbf n_2,\mathbf
n_3\rangle}{\sqrt 3} \;, \label{GHZW}
\end{align}
where $\{\mathbf n_i,-\mathbf n_i\}$ denotes an arbitrary local
orthonormal basis in the Hilbert space of qubit $i$. Note, that
for the original $GHZ$ state the phase $\alpha = 0$ and for the
original $W$ state the phases $\beta=\gamma=0$, and the
corresponding parameters $\Lambda$ appearing in the witness
operator~(\ref{wit}) are $1/2$ and $2/3$, respectively
\cite{Acin}. The states~(\ref{GHZW}), however, can be transferred
to the original ones by local unitary operations, which leave the
parameters $\Lambda$ unchanged, that is we have the witness
operators
\begin{align}
W_{GHZ(\alpha)} &= \frac{1}{2} \mathbb I_8
-|GHZ(\alpha)\rangle\langle GHZ(\alpha)|
\nonumber \\
W_{W(\beta,\gamma)} &= \frac{2}{3} \mathbb I_8
-|W(\beta,\gamma)\rangle\langle W(\beta\gamma)| \label{witGHZW}
\end{align}
for the $GHZ$- and $W$-like states, respectively.

Lunkes et al.~\cite{LBV} applying these witness operators for the
special case $\alpha=\beta=\gamma=0$ and $|\mathbf
n_i\rangle=|\uparrow\rangle$, $i=1,2,3$ established that neither
$\mathrm{ Tr}(\rho_3 W_{GHZ})$ nor $\mathrm{Tr}(\rho_3 W_W)$ can
become negative in the permitted range of parameters $p_{ij}$,
$ij=12,13,23$, hence no GTE could be revealed in the three-spin
reduced density matrix $\rho_3$ by the application of these
witnesses.

We confirm and extend this result by generalizing the $GHZ$ and
$W$ witnesses to the form~(\ref{witGHZW}) with the corresponding
values $\alpha,\beta,\gamma \in [0,2\pi]$ and with the arbitrary
local bases $|\mathbf n_i\rangle = \cos
(\theta_i/2)|\uparrow\rangle +
e^{i\phi_i}\sin(\theta_i/2)|\downarrow\rangle,\; i=1,2,3$. Then
the trace of $\rho_3 \Pi$, where $\Pi$ denotes either
$W_{GHZ(\alpha)}$ or $W_{W(\beta,\gamma)}$ and $\rho_3$ is the
state~(\ref{rho3}), is a linear combination of the trigonometric
functions cosine/sine with arguments $\theta_i,\phi_i,\;i=1,2,3$.
Owing to convexity arguments $\mathrm{Tr}(\rho_3 \Pi)$ can be
extremal only if $p_{ij}\in\{+1,-1\},\; ij=12,13,23$ in the
permitted range~(\ref{pijbound}) and $\theta_i,\phi_i \in \{0,
\pi/2, \pi, 3\pi/2, 2\pi\}, \; i=1,2,3$. Moreover, we may fix
three parameters, e.g., $\phi_1=\phi_2=0$ and $\theta_1=0$, owing
to the invariance of the state $\rho_3$ under the collective
$SU(2)$ rotation. Considering all the possible combinations of the
remaining parameters $p_{ij}$, $ij=12,13,23$ and
$\theta_2,\theta_3,\phi_3$ from the set above, we found that
$\mathrm{Tr}(\rho_3 \Pi)\geq 0$ is always true for the witness
operators $\Pi$ in (\ref{witGHZW}). Therefore, genuine tripartite
entanglement could not be witnessed by the general $GHZ$-type and
$W$-type witnesses~(\ref{witGHZW}), as well.

Naturally, we may ask whether there exist other witnesses over the
$GHZ/W$-types in~(\ref{witGHZW}) which are better suited for
detecting GTE in the state $\rho_3$. Indeed, in Ref.~\cite{CC05}
it has been shown that there are several genuine triparite
entangled states which are not witnessed by the
operators~(\ref{witGHZW}). Note, that nonlinear entanglement
witnesses \cite{GL06} may show improvement with respect to linear
witnesses in the multipartite case as well. However, if we stick
to linear functionals we are even able to reveal GTE in the state
$\rho_3$, as it will be discussed in the next subsection.

\subsection{The witness operator of G\"uhne et al.}
\label{subsec:witness}

G\"uhne et al.~\cite{GTB} showed that the internal energy is a
good indicator of genuine tripartite entanglement in macroscopic
spin systems. The idea was to write the internal energy in terms
of the mean value of the observables  $W_{ijk}=
\vec{\sigma}^i\cdot\vec{\sigma}^j + \vec{\sigma}^j \cdot
\vec{\sigma}^k$, where
$\vec{\sigma}^i=(\sigma_x^i,\sigma_y^i,\sigma_z^i)$ is the vector
of Pauli spin operators associated with the qubit $i$, and the
absolute value of $\langle W_{ijk} \rangle$ has been shown to be a
witness itself, capable to detect GTE. Namely, it has been proven
\cite{GTB} that if the inequality
\begin{equation}
|\langle W_{ijk} \rangle|> 1+\sqrt 5\simeq 3.236 \label{Wijk}
\end{equation}
holds, the qubits $i,j,k$ are genuinely tripartite entangled.

Let us use this inequality~(\ref{Wijk}) in order to reveal GTE in
the degenerate Fermi gas among the spins of the three fermions
$i,j,k$. Plugging the state $\rho_3$ in (\ref{rho3}) into the
expectation $\langle W_{ijk}\rangle =\mathrm{Tr}(\rho_3 W_{ijk})$
one has $\langle W_{ijk} \rangle=3(p_{ij}+p_{jk})$, which by
substitution back into (\ref{Wijk}) gives the condition
\begin{equation}
|\langle W_{ijk} \rangle|:= 3|p_{ij}+p_{jk}|> 1+\sqrt 5
\label{Wijk3}
\end{equation}
for the existence of genuine tripartite entanglement in the
degenerate Fermi gas. Next we discuss from the viewpoint of
witnessed GTE by the mean of this condition two different
three-fermion configurations:

(a) Consider the case investigated before in
Section~\ref{subsec:param}, that three particles lie evenly spaced
on a line close to each other. Choosing $ijk=123$ and recalling
$p_{12}=p_{23}=2/3$ from (\ref{plimit}), by the virtue of
(\ref{Wijk3}), $|\langle W_{123} \rangle|=4 > 1+\sqrt 5$, hence
the three-fermion state $\rho_3$ corresponding to this arrangement
of particles is genuine tripartite entangled.

(b) In this case the particles are separated from each other by
equal distances, i.e., the particles are put on the vertices of an
equilateral triangle. Owing to three-fold symmetry of this
configuration the state $\rho_3$ contains an equal mixture of
maximally entangled states $|\Psi^-\rangle$, that is, all
$p_{ij},\;ij=12,13,23$ in (\ref{rho3}) must have the same value.
Further, considering the constraint
$|p|=|p_{12}+p_{13}+p_{23}|\leq 1$ in (\ref{p}) and also owing to
the left-hand side of (\ref{Wijk3}) we have $|\langle
W_{123}\rangle|=|\langle W_{231} \rangle|=|\langle
W_{132}\rangle|=2|p|\leq 2$ implying that in this case all
possible $|\langle W_{ijk}\rangle|$ (with different permutations
of $ijk$) are smaller than the bound $1+\sqrt 5$. Consequently, no
GTE can be revealed by the witness~(\ref{Wijk}) of G\"uhne et al.,
no matter how far the fermions are separated from each other. It
is reasonable to think that there is indeed no GTE associated with
this highly symmetrical configuration, as it has been argued in
Ref.~\cite{LBV} by attributing it to the Pauli principle. In the
next section we construct from the observable $W_{ijk}$ a witness
operator $\tilde{W}_{ijk}$, which has a maximum eigenvalue smaller
than unity ($\tilde{W}_{ijk}\leq \mathbb I_8$) and with the aid of
it a lower bound is given for the amount of GTE in the state
$\rho_3$ quantified by an entanglement monotone, the generalized
robustness $E_R$.

\section{Deriving a lower bound to the generalized robustness $E_R$}
\label{sec:er}

Up to this point the observable $W_{ijk}$ was applied for
witnessing genuine tripartite entanglement. On the other, they are
also good for quantifying it \cite{Bra05} (see~\cite{RGFGC} as an
application to a magnetic material). The maximum eigenvalue of the
operator $-W_{ijk}$ is 4, thus considering~(\ref{Wijk}) we may
construct the following witness operator,
\begin{equation}
\tilde{ W}_{ijk} = \frac{(1+\sqrt 5)\mathbb I_8 - W_{ijk}}{5+\sqrt
5}\;,\label{tildeWijk}
\end{equation}
whose negative mean value $\mathrm{Tr}(\rho \tilde{W}_{ijk})$
guarantees that the three-qubit state $\rho$ is genuine tripartite
entangled. Further the witness operator is normalized so that
$\tilde{W}_{ijk}\leq \mathbb I_8$. It is apparent that provided
the mean value $\langle W_{ijk} \rangle \geq 0$ for state $\rho_3$
(i.e., $p_{ij}+p_{jk} \geq 0$ according to calculations in
Sec.~\ref{subsec:witness}), the witness operator~(\ref{tildeWijk})
is just as powerful to detect GTE associated with state $\rho_3$
as $|\langle W_{ijk}\rangle|$ in the inequality~(\ref{Wijk}).

The generalized robustness $E_R$ as a GTE measure quantifies how
robust the genuine tripartite entangled state $\rho$ is under the
influence of noise, and also has a geometrical meaning measuring
the distance of $\rho$ from the biseparable set $B$ \cite{CBC}.
According to Ref.~\cite{Bra05} $E_R$ can be expressed in a
Lagrange dual representation
\begin{equation}
E_R(\rho) = \max\{0,-\min_{\Pi\in M}\mathrm{Tr}(\rho\Pi)\}\;,
\label{witent}
\end{equation}
where the set $M$ is given by the restriction $\Pi\leq \mathbb
{I}_8$ for the GTE witnesses $\Pi$.

Since $\tilde{W}_{ijk}$ with any permutation of $ijk$ defines a
valid GTE witness with maximum eigenvalue smaller than unity, we
are able to develop the lower bound
\begin{equation}
E_{R,\min}(\rho) = \max\{0,-\min_{ijk \in \{123,231,132 \}}
 \mathrm{Tr}(\rho \tilde{W}_{ijk})\}
\label{boundER}
\end{equation}
to the generalized robustness~(\ref{witent}) of an arbitrary state
$\rho$ on qubits $123$, as it is discussed in
Refs.~\cite{CC06,EBA}. This lower bound by
applying~(\ref{tildeWijk}) for the particular state $\rho_3$ in
(\ref{rho3}) reads as
\begin{equation}
E_{R,\min}(\rho_3) =
\max_{ijk\in\{123,231,132\}}\{0,\frac{3(p_{ij}+p_{jk})-1-\sqrt
5}{5+\sqrt 5} \}\;. \label{boundER3}
\end{equation}

In the next section this formula will be applied to give
explicitly a lower bound to the generalized robustness $E_R$ of
GTE for various configurations of three fermions, associated with
the reduced state $\rho_3$ of the degenerate Fermi gas.

\section{Numerical calculations to the lower bound of $E_R$}
\label{sec:num}

\subsection{Fermion moving on a straight line}
\label{subsec:line}

Now we concentrate on two different kinds of arrangements of the
three fermions in the 3D degenerate Fermi gas, which
configurations were also investigated in Ref.~\cite{LBV} from the
viewpoint of bipartite entanglement shared between two arbitrary
groups of three fermions.

\begin{figure}
\centerline{\epsfxsize 2.8in \epsffile{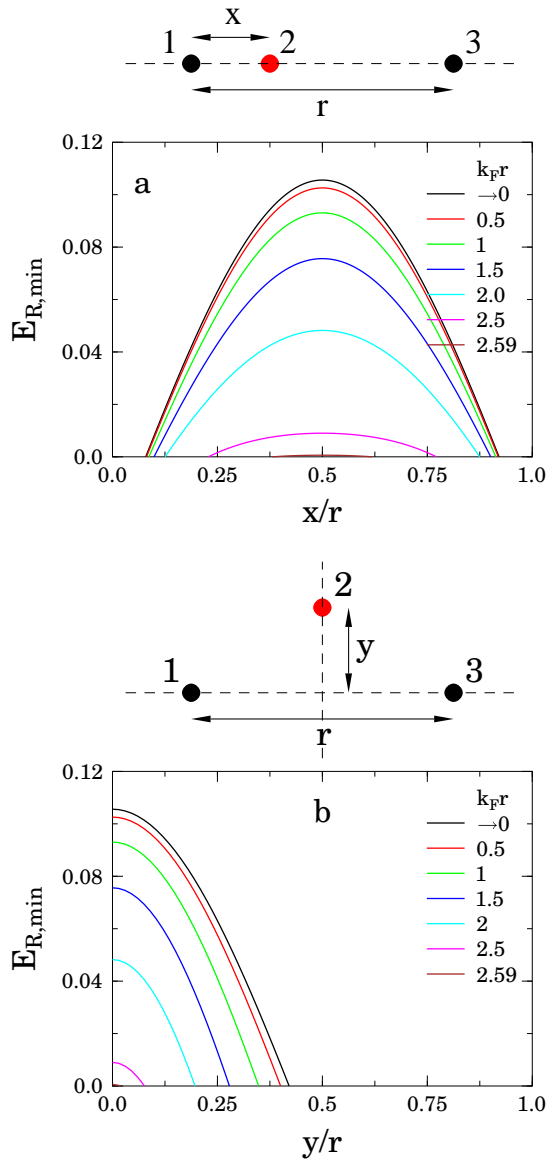}}
\caption{(color online) Lower bound to the generalized robustness
$E_R$ of genuine tripartite entanglement shared by three fermions
is plotted for the 3D Fermi gas for the cases (a) fermion $2$ is
moved away from the position of fermion $1$ toward the position of
fermion $3$, (b) fermion $2$ is moved away from the midpoint of
the line connecting fermion $1$ and fermion $3$ normal to this
line (see the respective schematic geometrical pictures). The
curves are plotted in both cases for the following dimensionless
distances between fermion $1$ and fermion $3$: $k_Fr\rightarrow0$
(displayed in black) and $k_Fr=0.5,1,1.5,2,2.5,2.59$ (displayed in
color).} \label{fig-line}
\end{figure}

(a) In the first instance a collinear arrangement is considered,
namely we put three fermions on a straight line numbering them in
the order $1,2$, and $3$. The distance between particles $1$ and
$3$ is $r$, and the intermediate particle $2$ is by a distance of
$x$ away from particle $1$ (shown by the geometrical picture of
Fig.~\ref{fig-line}.(a)). In Fig.~\ref{fig-line}.(a), the lower
bound to GTE quantified by $E_R$ is plotted in the $3D$ Fermi gas
according to the formula~(\ref{boundER3}) in the function of $x/r$
for different values $k_Fr$ of the external fermions. The
calculations can be in general performed only numerically, however
for the limiting value $k_Fr\rightarrow 0$ one obtains
\begin{equation}
\max_{ijk\in 123, 231, 132}{\{p_{ij}+p_{jk}\}} = p_{12}+p_{23} =
\frac{1}{1-x/r+(x/r)^2}\;.\label{p123x}
\end{equation}
Substitution of this expression into (\ref{boundER3}) gives
analytically the curve corresponding to $k_Fr\rightarrow 0$. Note
that formula~(\ref{p123x}) holds true independently of the
dimensionality of the Fermi gas (i.e., both for the 2D and 3D
cases). The curves produced in Fig.~\ref{fig-line}.(a) exhibit two
essential features: For any given value of $k_Fr$, the maximum of
$E_{R,\min}$ is achieved by the symmetrical configuration (i.e.,
particle $2$ is located at the midpoint of the line connecting
particles $1$ and $3$), still presenting GTE in the system by the
dimensionless distance $k_Fr=2.59$. On the other, when fermion $2$
starting from this midpoint is moved toward fermion $1$ in the
case $k_Fr\rightarrow 0$, the curve falls off to zero by the value
$x/r =1/2 (1 - \sqrt{3(\sqrt 5 -2)}) \simeq 0.08 $. That is, if
fermions $1$ and $3$ are a distance $r$ away from each other and
fermion $2$ becomes closer than $x \simeq 0.08r$ to fermion $1$
(or to fermion $3$ in the symmetrically equivalent situation)
formula~(\ref{boundER3}) does not indicate GTE among the three
fermions. This result fits to the monogamy property of
entanglement \cite{CKW}, as in the case $x/r\rightarrow 0$ fermion
$2$ becomes maximally entangled with fermion $1$, excluding the
existence of any higher order entanglement in the system.

(b) Now let the three particles lie on the vertices of an
isosceles triangle fermions $1$ and $3$ forming its base with
length $r$, and fermion $2$ is positioned by a distance of $y$
from the midpoint of the base as it is illustrated in the
geometrical part of Fig.~\ref{fig-line}.(b). The curves
$E_{R,\min}$ in Fig.~\ref{fig-line}.(b) are plotted against $y/r$
for different values of $k_Fr$. As it can be observed, all the
curves $E_{R,\min}$ plotted are monotonically decreasing in the
function of the ratio $y/r$ for any given $k_Fr$. Similarly to
case (a) the curve corresponding to $k_Fr\rightarrow 0$ can be
treated analytically, and one obtains vanishing GTE beyond the
value $y/r =1/2 (\sqrt{3(\sqrt 5 -2)}) \simeq 0.42$ (both in the
2D and 3D Fermi gases). This supports the result of case (b) in
Sec.~\ref{subsec:witness} that three fermions located at the
vertices of an equilateral triangle (where $y/r = \sqrt 3/2 \simeq
0.866$) are not genuine tripartite entangled independent of the
separation distance $r$. It is also apparent from
Fig.~\ref{fig-line}.(b) that for a fixed ratio $y/r$, $E_{R,\min}$
is maximal when $k_Fr\rightarrow 0$, for in this case the
antisymmetrization effect between the three particular fermions
and the rest of the fermions (which reduces the amount of quantum
correlations shared among the three fermions) becomes negligible.

\subsection{Fermion moving in a plane}
\label{subsec:plane}

\begin{figure}
\centerline{\epsfxsize 3.2in \epsffile{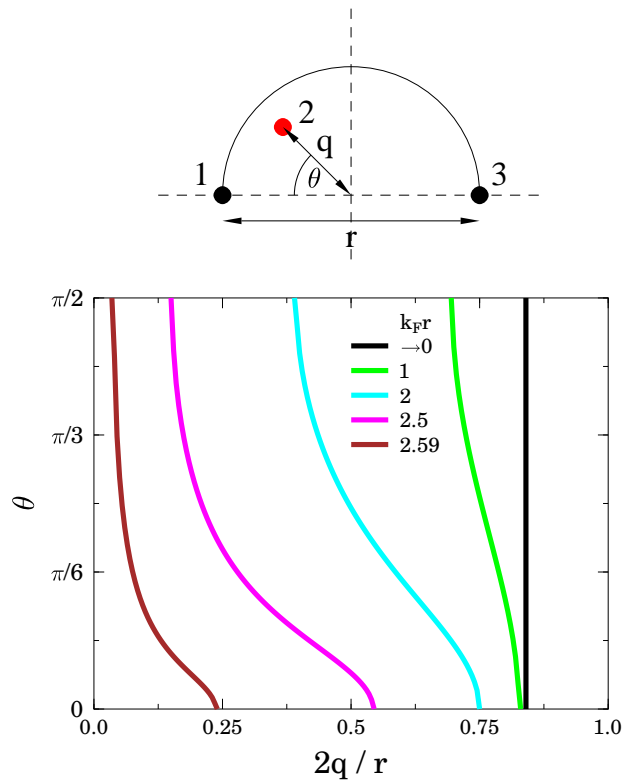}}
\caption{(color online) Witnessed genuine tripartite entanglement
among three fermions is plotted for the 3D Fermi gas for the
configuration where fermion $2$ (with polar coordinates
($\theta,q$)) is allowed to move in a disk of radius $r/2$
centered at the midpoint between fermion $1$ and fermion $3$, as
shown by the geometrical picture. The polar plot represents curves
with black, green, cyan, magenta and brown colors, which separate
the regions associated with GTE (left-hand side) from regions
without having GTE (right-hand side) for values of the
dimensionless distances $k_Fr\rightarrow 0$, $k_Fr=1,2,2.5,2.59$,
respectively.} \label{fig-plane}
\end{figure}

We now turn to the situation (pictured in the geometrical part of
Fig.~\ref{fig-plane}) when fermion $2$ is allowed to move in the
two-dimensional plane given by polar coordinates ($\theta,q$) with
an origin at the midpoint of the line of length $r$ connecting
fermion $1$ and fermion $3$. Let us restrict fermion $2$ to be
positioned within the circle with radius $r/2$. Recalling the
definition of the GTE distance from Sec.~\ref{sec:intro}, in this
case the GTE distance is equal to the maximum separation length
$r$ between the two external fermions, below which the three
fermions are genuine tripartite entangled.

In what follows, we inquire the shape of region within particle
$2$ could be located so that fermions $1,2$, and $3$ by a fixed
relative distance $r$ would be genuine tripartite entangled.
Further, owing to the symmetry of the configuration it suffices to
study the interval $\theta \in [0,\pi/2]$, i.e., particle $2$ is
restricted to be situated on a quarter-disk of radius $r/2$. The
polar plot of Fig.~\ref{fig-plane} represents curves which
distinguish regions with (left side) and without (right side)
witnessed GTE in the system, by the following values of the
dimensionless distances $k_Fr\rightarrow 0$ and
$k_Fr=1,2,2.5,2.59$. The witnessed GTE (i.e., to decide whether
$E_{R,\min}(\rho_3)>0$) corresponding to these curves for the
various values of $k_Fr$ was calculated according to the
formula~(\ref{boundER3}).

The case $k_Fr\rightarrow 0$ can be treated analytically yielding
the result (as it can be read off from Fig.~\ref{fig-plane}) that
by $k_Fr\rightarrow 0$ GTE is present in the system provided
fermion $2$ is located inside a disk with radius $q\simeq 0.42r$
(generalizing the results $x\simeq 0.08r$ and $y\simeq 0.42r$ in
the limit $k_Fr\rightarrow 0$ obtained in Sec.~\ref{subsec:line}).
Now one can see from the shape of the polar curves that for
greater $k_Fr$ the corresponding disk associated with GTE squeezes
toward the axis of particles $1$ and $3$, eventually contracting
on the origin $q=0$. It is noted that the same behavior would have
been observed for the case of 2D Fermi gas, as well. This implies
that if in this particular case we were searching for the GTE
distance (i.e., the maximum separation $r$, below which GTE
exists) then we should focus on the case where particle $2$ is
located at the midpoint between particles $1$ and $3$. This task
will be performed in the sequel both for the 2D and 3D degenerate
Fermi gases.

\section{An upper and a lower bound to the GTE distance in the 2D
and 3D degenerate Fermi gases} \label{sec:gte}

\subsection{Lower bound}
\label{subsec:lower}

In the present section the three-fermion configuration is
investigated, where the particles $1,2$ and $3$ are positioned
evenly spaced on a straight line, with a relative distance $r$
between the external particles $1$ and $3$. We develop both for
the 2D and 3D Fermi gases an upper and a lower bound to the
relative distance $r$, beyond which GTE disappears.

\begin{figure}
\centerline{\epsfxsize 2.8in \epsffile{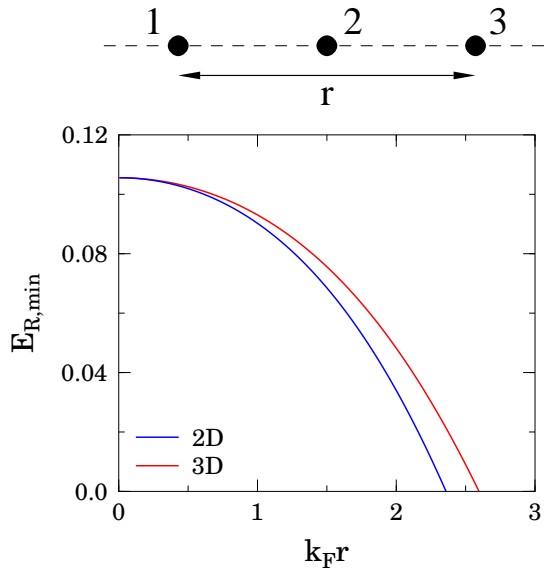}}
\caption{(color online) Lower bound to the generalized robustness
$E_R$ of genuine tripartite entanglement is plotted both for the
two- and three-dimensional degenerate Fermi gases in the function
of the dimensionless distance $k_Fr$, where $r$ is the relative
distance between the positions of fermion $1$ and fermion $3$, and
fermion $2$ is located at the midpoint, as shown by the upper part
of the figure.} \label{fig-distance}
\end{figure}

Let us first calculate numerically and plot the lower bound to the
generalized robustness of GTE defined by~(\ref{boundER3}) in the
function of $k_Fr$. In Fig.~\ref{fig-distance} the respective
curves are plotted both for the two- and three-dimensional
degenerate Fermi gases, and numerics shows that
$E_{R,\min}(\rho_3)$ vanishes beyond, i.e., the lower bound to the
GTE distance is
\begin{align}
r_{\min}^{\mathrm{2D}} &=2.3588/k_F\;,  \nonumber \\
r_{\min}^{\mathrm{3D}} &=2.5964/k_F\;, \label{rmin}
\end{align}
respectively. Let us compare these values (\ref{rmin}) with the
bipartite entanglement distance (i.e., the maximal separation
distance between two entangled fermions), which are slightly
smaller, and are given explicitly by the values \cite{LBV}
$1.6163/k_F$ and $1.8148/k_F$ for the 2D and 3D Fermi gases,
respectively.

For the 3D Fermi gas the $p_{ij}$ parameters of the state $\rho_3$
corresponding to $r_{\min}^{\mathrm{3D}}$ are
\begin{equation}
p_{12}=p_{23}=0.539345 ,\;\; p_{13}=-0.160702\;. \label{p123min}
\end{equation}
We can also observe in Fig.~\ref{fig-distance}, as one may expect,
that the curves are monotonically decreasing, such as in the
bipartite case for the entanglement measure negativity \cite{LBV}.

\subsection{Upper bound}
\label{subsec:upper}

We continue with studying the particular three-fermion
configuration discussed in the previous subsection in order to
establish an upper bound to the GTE distance, beside the lower
bound already obtained. Exploiting the mirror symmetry of the
configuration, for any given distance $r$ between the positions of
particle $1$ and particle $3$ we have $p_{12}=p_{23}$, thus in
this case parameters $r_k$ defined by (\ref{rk}) in
Section~\ref{subsec:3spin} can be expressed through $p_{12}$ and
$p_{13}$ alone, and we obtain the relation
\begin{equation}
r_2=\sqrt 3 r_1\;. \label{rsym}
\end{equation}

\begin{figure}
\centerline{\epsfxsize 3.0in \epsffile{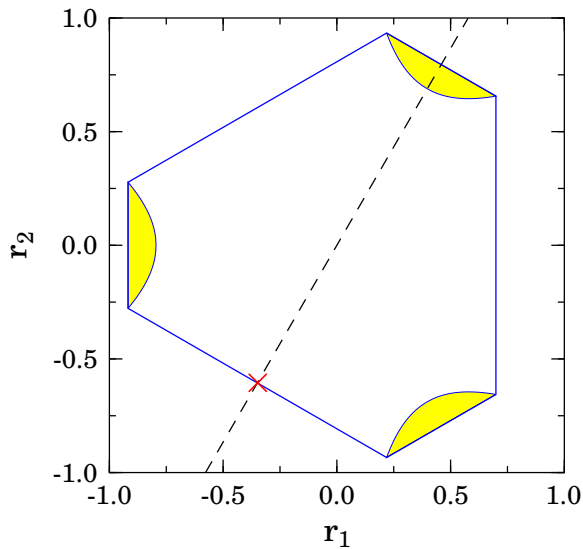}}
\caption{(color online). The three yellow shaded half-disks
represent areas corresponding to biseparable states with respect
to the partitions $1|23$, $12|3$ and $13|2$ on the $(r_1,r_2)$
plane by the section $r_+=0.041$ and $r_3=0$ for the 3D Fermi gas.
The solid blue straight lines bound the set of biseparable states
$B$ at this particular section. The equation $r_2=\sqrt 3 r_1$ is
represented by the dashed line, and the red cross stands for the
point which is located on this line with coordinate $r_1=-0.35$.}
\label{fig-polygon}
\end{figure}

On the other, by applying Theorem~7 of Eggeling and Werner
\cite{EW} it is asserted that a three-qubit state is biseparable
with respect to the partition $1|23$ if the following inequalities
are satisfied:
\begin{eqnarray}
-1< r_1-2r_+ < 0\;, \nonumber \\
3r_2^2 +3r_3^2+(1-3r_+)^2 \leq (r_1-2r_+)^2\;. \label{EWineq}
\end{eqnarray}
Now let us calculate the boundary of the area in the plane
$(r_1,r_2)$ described by these inequalities~(\ref{EWineq}) by
$r_+=0.041$ and $r_3=0$, which parameters correspond to the
symmetrical configuration of separation $r_{\min}^{\mathrm{3D}}$
between fermion $1$ and fermion $3$ with
parameters~(\ref{p123min}). The solution of the
inequality~(\ref{EWineq}) by $r_+=0.041$, $r_3=0$ corresponds to
the leftmost yellow shaded semi-disk in Fig.~\ref{fig-polygon},
representing states in the section $r_+=0.041$ and $r_3=0$ that
are separable with respect to the partition $1|23$. The other two
disks (representing biseparable states with respect to partitions
$12|3$, $13|2$) can be obtained through $\pm 2\pi/3$ rotations
around the origin of the plane $(r_1,r_2)$ \cite{TA} owing to the
permutation symmetry of the three subsystems. The boundary of the
convex hull of these semi-disks are indicated by solid blue line
segments in Fig.~\ref{fig-polygon}. All the three-qubit states in
the section $r_+=0.041$ and $r_3=0$ which lie inside this polygon
are in the class of biseparable states $B$. However, since this
area corrresponds to a section, biseparable states in this section
may exist outside the polygon as well.

Next let us determine the explicit position of the point
corresponding to the mirror-symmetrical configuration with
separation $r_{\min}^{\mathrm{3D}}$ between the two external
fermions. In this particular symmetrical configuration according
to~(\ref{rsym}) the ratio $r_2/r_1=\sqrt 3$ and we also have
$r_1=(p_{13}-p_{12})/2$. The above ratio has been displayed in
Fig.~\ref{fig-polygon} by a dashed line and the point on this line
with coordinate $r_1 = -0.35$ (obtained by plugging the
values~(\ref{p123min}) into the above formula for $r_1$)
corresponding to the separation $r_{\min}^{\mathrm{3D}}$ with
$E_{R,\min}=0$ has been designated by the red cross marker.
Numerics shows, that this point lies outside the solid blue
polygon, as it ought to owing to $E_R\geq 0$ associated with the
point. By inspection, on the other, this point is very close to
the border of the polygon.

Indeed, explicit numerical calculations yield that the distance
$r$ between the two outer fermions corresponding to the border of
the polygon is $r_{\max}^{\mathrm{3D}}=2.5988/k_F$. This value was
obtained by tuning the value of $r_+$ from $0.041$ up to $\simeq
0.0415$ by the mean of increasing the separation $r$ starting from
the value $r_{\min}^{\mathrm{3D}}$ so that by
$r_{\max}^{\mathrm{3D}}$ the point represented by the the cross
marker would lie just on the edge of the polygon. However, this
distance $r_{\max}^{\mathrm{3D}}=2.5988/k_F$ is just an upper
bound to the GTE distance for the 3D Fermi gas. On the other,
similar evaluations give the value
$r_{\max}^{\mathrm{2D}}=2.3599/k_F$ for the 2D Fermi gas.
Comparing these values with the ones in~(\ref{rmin}) corresponding
to the lower bound, it shows that the upper and lower bounds to
the GTE distance are indeed very close to each other both for the
2D and 3D Fermi gases. Hence, this implies that the witness of
G\"uhne et al.~\cite{GTB} in our particular problem ought to be
close to an optimal one.

\section{Discussion}
\label{sec:disc}

Previous works (e.g., \cite{Ved03,LBV}) explored that bipartite
entanglement exists within the order of the Fermi wavelength
$1/k_F$ at zero temperature in the non-interacting Fermi gas and
may even persist for nonzero temperatures (e.g., \cite{LBVa,OK}).
Since the system consists of non-interacting fermions,
entanglement is purely due to particle statistics and not to any
physical interaction between the particles. In the present work we
found the result as an extension of the formerly studied bipartite
case that particle statistics is capable to generate genuine
tripartite entanglement (GTE) as well. Furthermore, it has been
found that the diameter of the three-fermion configuration wherein
GTE is present (a lower bound to the maximum diameter is given
explicitly by~(\ref{rmin})) is comparable with the maximum
relative distance between two entangled fermions, both in the 2D
and 3D Fermi gases. Looking at higher order entanglement as a
useful resource, the presence of GTE in Fermi systems would be
promising to allow for performing new quantum information
processing tasks, exemplified by the $GHZ$ paradox \cite{GHSZ}.
However, in order to do so, the amount of entanglement stored by
three fermions should be somehow extracted from the system. In the
present article, though, we did not consider this problem some
explicit schemes has been put forward recently in the bipartite
setting \cite{DDW, Cav06, CV}, some of which might be extended to
the tripartite setting as well.

Namely, in Ref.~\cite{DDW} it has been shown that bipartite
entanglement can exist between non-interacting fermions on a
lattice and can extend over multiple lattice sites even if the
entanglement is quantified by the most restrictive measure, the
entanglement of particles \cite{WV}. Further, considering that in
the continuum limit the entanglement of particles corresponds to
the entanglement in the spin reduced density matrix \cite{DDW}, by
continuity arguments genuine tripartite entanglement, quantified
by the measure entanglement of particles, should exist in the
lattice system as well. Thus, in the near future optical lattice
implementations may offer a simulation technics to observe the
phenomenon of genuine tripartite entanglement among
non-interacting fermions in a lattice.

On the other, in the continuum limit the extraction of genuine
tripartite entangled particles seems to be a more difficult
problem: As it has been shown \cite{Ved03} the entanglement
distance between two fermions is inversely proportional to the
Fermi momentum $k_F$, and $k_F^3$ in turn is proportional to the
density of particles. In the case of conduction electrons in a
usual metal the density is very large indicating an entanglement
distance of the order of a few angstroms. This failure might be
avoided by using 2D electron gas formed in GaAs heterostructure,
where the entanglement distance is in the order of hundred
angstroms \cite{OK} or using stored ultra-cold neutrons in a
carefully devised experiment \cite{CV}. Also, note the intriguing
proposal, exploiting decoherence effects to extract bipartite
entanglement created merely by particle statistics from
semiconductor quantum wells \cite{Cav06}. Although the GTE
distance in (\ref{rmin}) is comparable (even greater) than the
bipartite entanglement distance both for the 2D and 3D Fermi
gases, technically these proposals appear to be very demanding
when applied to the three-party setting.

Finally, we would like to mention interesting future directions as
a continuation of the present work. One could for example apply
the same methods as in the present article for determining GTE
distance in Fermi gases trapped in a harmonic trap \cite{Yi} or
considering GTE not only in spin, but in other internal degrees of
freedom as well \cite{CWZ}. Also the possible existence of genuine
multipartite entanglement beyond the three-party scenario remains
to be explored.

\acknowledgments

This work was supported by the Grant \"Oveges of the National
Office for Research and Technology.

\end{document}